\documentclass{elsart}

\usepackage{graphicx,amssymb,amsmath,float}
\begin{document}

\begin{frontmatter}

\title{Volatility  in atmospheric temperature variability}
\author[adr1,adr2]{R. B. Govindan \corauthref{cor1}}
,
\author[adr1]{Armin Bunde},
\author[adr2]{Shlomo Havlin}
\corauth[cor1]{Corresponding author. E-mail:
govindan@mpipks-dresden.mpg.de\\{{\bf Present Address:} 
\\Max Planck Institut f\"{u}r Physik komplexer Systeme,\\ N\"{o}thnitzer Strasse
38, \\D-01187 Dresden, \\Germany} }
\address[adr1]{Institut f\"{u}r Theoretische Physik III, Justus-Liebig-Universit\"{a}t 
Giessen, Heinrich-Buff-Ring 16, 35392 Giessen, Germany}
\address[adr2]{Minerva Center and Department of Physics, Bar-Ilan University, Ramat-Gan 
52900, Israel}
\begin{abstract}
Using detrended fluctuation analysis (DFA), we study the scaling
properties of the volatility time series $V_i=\vert T_{i+1}-T_i\vert$ of daily 
temperatures $T_i$ for ten  
chosen sites around the  
globe. We find that the volatility is long range power-law correlated with an  exponent $\gamma$ close to 0.8 for all
 sites considered here. We use
this result to test the scaling performance of several state-of-the art  global climate models and find that
the models do not reproduce the observed scaling behavior.  
\end{abstract}
\end{frontmatter}

\section*{\bf 1. Introduction}
Indications of weather persistence over months and seasons are known
\cite{sck}.   
This long term persistence in the atmospheric temperature
variability is 
analysed and quantified by detrended fluctuation analysis (DFA) and wavelet
transform techniques with a fluctuation exponent of $\alpha \sim 0.65 $, which is independent of the
location of the site \cite{1}.  Also, indications of persistence are
confirmed through
power spectral analysis \cite{2,3}. Presence of such a universal persistence
law indicates that the processes governing the atmospheric dynamics at
different climatological zones are based on similar principles
\cite{1}. Here, we are not interested in the temperature fluctuations
around the seasonal trend, but in the magnitude of temperature changes between
successive days, 
i.e the temperature volatility.  
Volatility is the concept often used in {\it econophysics} to indicate
the fluctuations in price changes \cite{pa299}. The market is said to
be more volatile if the fluctuations in price changes are high \cite{pre60}.
In recent years this concept has been applied to cardiac system and found that
cardiac volatility series is long range correlated \cite{prl86}.

Here we study the
volatility of
 atmospheric temperature data obtained 
from 10 randomly chosen meteorological stations in  
Europe, North America, Asia, Russia and Australia, from various climatological
zones. Correlations in the volatility series give information about
persistence in the changes. If, for example, the change in the
temperature between two successive days is small there is a high
tendency that the change remains similar for the next consecutive
days. Here, we study long-term temperature records (typically 100
years).  We use detrended fluctuation analysis (DFA) up to the order
of five \cite{5,AB,JWK}, which
systematically eliminates higher order trends and reveals the
correlations present in 
the highly non-stationary data. Our analysis shows
that (i) the
persistence, characterised by the correlation $C(s)$ of the
volatility series separated by $s$ days, follows a power law,
$C(s) \sim s^{-\gamma}$, with roughly the same exponent $\gamma \cong
0.8$ for all stations considered, and that (ii) the range of this
universal persistence law seems exceed one decade.

\section*{\bf 2. Methodology of Scaling analysis}
We have performed the scaling analysis of volatility series for the records of the maximum
daily temperature $T_i$ of the following weather stations: Prague (218 yr),
Melbourne (136 yr), Luling (90 yr), Seoul (86 yr), Kasan (96 yr), Vancouver (93
yr), Tashkent (97 yr),  New York city (116 yr), Brookings (99 yr) and
St. Petersburg (111 yr). The numbers within (.) are the length of the
records. From the daily maximum temperature series, we construct the
volatility series $V_i$=$\vert T_{i+1}-T_i \vert$. Then we remove the 
climatological annual cycle \cite{6} from $V_i$ to obtain
$\Delta V_i = V_i-<V_i>$.      

Qualitatively, persistence is clearly seen (patches of high and low
volatility) in the plot of $
V_i $  as shown in Fig. 1a
for one year in Prague. Figure 1b shows the volatility
series $\Delta V_i$ around the mean. Figures 1e and 1f show the plot of $
V_i$ and $\Delta V_i$ respectively, obtained 
from the phase randomised surrogate
data (preserving the distribution) \cite{12} of the temperature increment
series $(T_{i+1}-T_i)$. Quantitatively,
persistence in $\Delta V_i$ can be 
characterised by the (auto)correlation function,
$$
C(s) \equiv \langle \Delta V_i~\Delta V_{i+s} \rangle =
\frac1{N-s} \sum\limits_{i=1}^{N-s} \Delta V_i~\Delta V_{i+s}, \eqno(1)
$$
where
$N$ is the number of days in the record. A direct calculation of $C(s)$
is hindered by the level of noise present in finite temperature
series, and by possible nonstationarities in the data.
Following \cite{7,8}, we do not calculate $C(s)$ directly, rather we study the
fluctuations in the volatility {\it profile} $Y_n = \sum_{i=1}^n \Delta V_i$.   
We divide the profile into non overlapping time windows of
length $s$ and determine the squared fluctuations of the profile (as
specified below) in each segment. The mean square fluctuations,
averaged over all segments of length $s$, are related to the
correlation function $C(s)$ (see below). For this study we employ a
hierarchy of methods that differ in the way the fluctuations are
measured and possible nonstationarities are eliminated (see e.g.,\cite{JWK,9}
for detailed description of the methods):

(i) In the simple fluctuation analysis (FA), we calculate the difference
of the profile at both ends of each segment. The square of this
difference represents the square of the fluctuations in each segment.

(ii) In the {\it first order} detrended fluctuation analysis, we
determine in each segment the best linear fit of the profile. 
The
variance of the profile from this straight line
represents the square of the fluctuations in each segment.

(iii) In general, in the $n$-th order DFA we determine in each segment
the best $n$-th order polynomial fit of the profile. The
variance of the profile from these best $n$-th order polynomials 
represents the square of the fluctuations in each segment. 
\begin{figure}   
\begin{center}
\includegraphics[width=5in,angle=0]{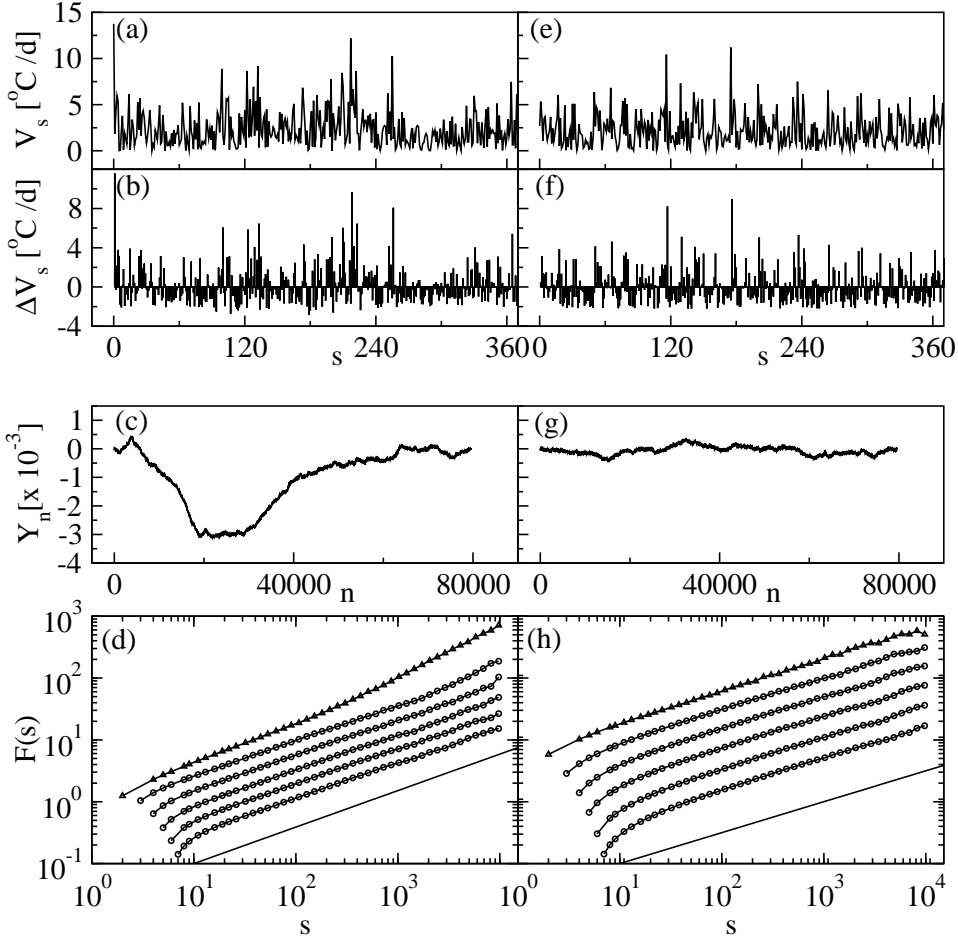} 
\vskip 0in
\caption 
{Temperature volatility series 
for Prague (1y), before 
removing the (climatological) annual cycle. (b) Same data as in (a)
but after removing the annual cycle. (c) Profile function of the
volatility series. (d) Fluctuation functions $F(s)$ obtained
from the profile 
function shown in (c). Curves from top to bottom represent $F(s)$
obtained from FA ($\triangle$) and DFA1-5 (o). Panels (e)-(h)
represent the same quantities as in (a)-(d) for phase randomised
surrogate data (preserving the distribution) of the temperature increment series $(T_{i+1}-T_i)$. Solid lines at the bottom 
of the panels (d) and (h) are lines with slopes 0.6
and 0.5 respectively. The unit of temperature volatility $V_i$ and its mean $\Delta
V_i$ is $^\circ$C/$d$. Scale of $F(s)$ shown in (d) and
(h) are arbitrary and the unit of $s$, $n$ and $d$ is a day.
}
\label{fig1}
\end{center}
\end{figure}
The fluctuation function $F(s)$ is the root mean square of the
fluctuations in all segments. For the relevant case of long-term
power-law correlations, $C(s) \sim s^{-\gamma}$, with $0<\gamma<1$, the
fluctuation function increases with $s$ according to a power law
\cite{10},
$$
F(s) \sim s^\alpha,~~~~~\alpha=1-\frac\gamma 2.  \eqno(2)
$$
For uncorrelated as well as short range correlated data, we have
$\alpha=\frac12$. For long range correlated data we have
$\alpha>\frac12$.

By definition, FA does not eliminate trends similar to the Hurst method
and the conventional power spectral methods \cite{11}. In contrast,
DFAn eliminates trends of order $n$ in the profile and $n-1$
in the original time series. Thus, from the comparison of
fluctuation functions $F(s)$ obtained from different methods one can
learn about long term correlations and types of trends, which cannot
be achieved by the conventional techniques.                    
\begin{figure}
\begin{center}
\includegraphics[width=5in,angle=0]{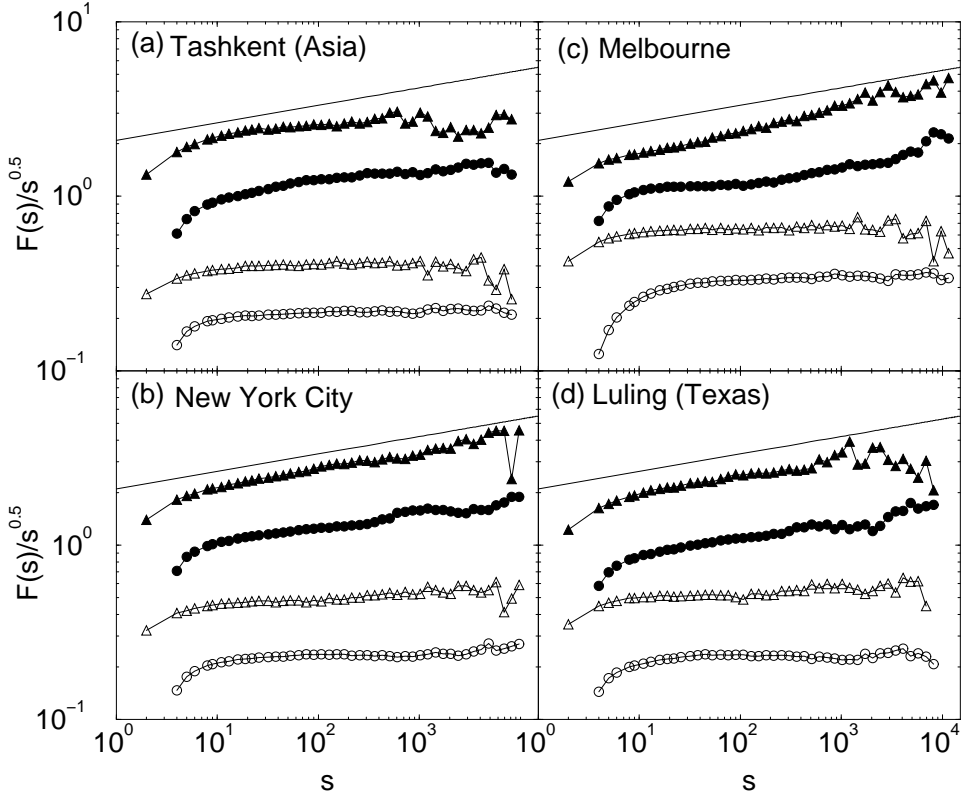} 
\caption 
{Fluctuation functions $F(s)$ obtained from temperature volatility
series for four different continental sites:
Tashkent (Asia), Melbourne (Australia), New York (US) and Luling
(US). Top two curves in each panel represent the
$F(s)$ obtained from FA ($\blacktriangle$) and DFA2 ($\bullet$) of
the volatility series. The third and fourth curves in each panel
represent 
the $F(s)$ obtained from FA ($\triangle$) and DFA2 (o)
of volatility series derived from the surrogate data. Scale of $F(s)$ is
arbitrary and the unit of {\it s} is a day. Solid lines shown at the top of
each 
panel is the line with slope 0.1 for comparison. Note that we plot
$F(s)/s^{0.5}$ so that a slope of 0.1 corresponds to 0.6.
}
\label{fig2}
\end{center}
\end{figure}  

\section*{3. Scaling analysis of Temperature volatility series}
We begin the analysis with the volatility series $\Delta
V_i$ for Prague which is the longest series (218 yr) in this
study. Figure 1c shows the profile, and Fig. 1d shows the fluctuation
functions obtained from FA ($\triangle$) and DFA1-5 (o).
 In the log-log
plot, all curves are approximately straight lines with a slope of
$\alpha$ close to 0.6. This result suggests that there
exists a long 
term persistence expressed by the power law decay of correlation
exponent $\gamma \cong 0.8$. There is slight upward bend in the $F(s)$
curve obtained from FA. This shows that there is a trend in the
volatility series, as observed in \cite{1} for the temperature
data. Hence, the effect of the city growth in Prague does not lead only
to an increase in the mean temperature of the city, but also to an
increase in the temperature volatility. This trend is not
removed in FA while DFA1 and subsequently 
higher orders of DFA have removed it. Figure 1g shows the
profile of the volatility series obtained from the surrogate data
of the temperature increment series. When compared with Fig. 1c, profile shows much less
fluctuations. Figure 1h 
shows the fluctuation functions $F(s)$ obtained from 
FA ($\triangle$) and DFA1-5 (o) for the profile shown in Fig. 1g. In the
log-log plot, the fluctuation functions $F(s)$  are straight lines
with a slope of 
$\alpha\sim\frac12$. This shows that the original data have nonlinear
features \cite{16,4}. 

Figures 2(a-d) show the results of volatility analysis 
for four different continental sites: Tashkent from Asia (97 yr), New York
city (116 yr), 
Melbourne (136 yr) and Luling from Texas (90 yr). The top two curves
in each panel are the fluctuation curves 
obtained from FA ($\blacktriangle$) and DFA 2 ($\bullet$) for
volatility series of temperature data while 
the third and fourth curves are those obtained from the same analyses
(FA and DFA2) 
of volatility series derived from the surrogate data. For the sake of clarity, 
we have divided the fluctuation function $F(s)$ by $s^{0.5}$.
The straight
line shown in each panel has a slope 0.1. The scaled fluctuation functions
$F(s)/s^{0.5}$ obtained from the 
original series $\Delta V_i$ (the first and second curves in all
panels shown in 
Fig. 2) are almost parallel to the line with slope 0.1.
The fluctuation functions $F(s)/s^{0.5}$ obtained from the surrogate data
(third 
and fourth curves in all panels shown in 
Fig. 2) are all parallel to x-axis.  

We obtained similar
results with exponents between 0.58 and 0.63 for all 10 sites
analysed. Thus, there exist a 
quite general persistence law in the volatility series obtained from
temperature data. The existence of such a general
persistence law in the
temperature volatility series indicates that the change in the temperature in
different climatic zones may be governed by the same basic principles,
leading to similar fluctuations in volatility series in different
places. 

\section*{4 Testing the volatility scaling performance of simulated temperature records}
We use this result to test the  
state-of-the art global climate models. In an earlier work, Govindan \etal \cite{RBG}, have shown that temperature data simulated by GCMs violate
the observed scaling 
behavior \cite{1}. Here we consider the scaling analysis of the
temperature volatility time series. We concentrate on four models,
for which data for all three scenarios 
viz. Control run (CR), (ii) greenhouse gas
forcing only (GHGF) and (iii) greenhouse gas plus aerosol forcing
(GHGPS), are available for the same simulation period. The GCMs are: (i)
CSIRO-Mk2 (Melbourne), (ii) ECHAM4/OPYC3 (Hamburg), (iii) CGCM1(Victoria,
Canada)  and (iv)
CCSR/NIES (Tokyo) (see \cite{15} for
details).   

In CR, the CO$_2$ content is fixed. In GHGF scenario,
one mainly considers the effect of greenhouse gases.
The amount of greenhouse gas forcing is taken from the historic data
until 1990 and then increased at a rate of 1\% per year. In GHGPS
scenario, the effect of aerosols (mainly sulphates) in the atmosphere
is taken into account which can mitigate and partially offset the
greenhouse warming. Although this scenario represents an
important step towards comprehensive climate simulation, the precise
role of aerosols in the mechanism of climate modelling is still unclear.
 
The temperature data simulated by these four models for three
different scenarios are available from the IPCC Data Distribution
Center \cite{15}. We extracted data for
nine representative sites around the globe (Prague, Melbourne,
Seoul,
 Vancouver, Kasan, Luling, Tashkent, New York and St. Petersburg). For
each model and each scenario, we selected the temperature records of
the four grid points closest to each site, and bilinearly interpolated
the data to the location of the site.

We generate volatility series from the monthly temperature
series by the same procedure (explained earlier) for daily temperature data.
We removed the seasonal periodic trends in $T_i$ 
(before constructing the increment series) and also in the volatility
series $V_i$. 
  Since the GCMs simulate temperature data in  monthly time scales, 
for comparison, we averaged the daily temperature
(observed) data (of
the nine climatological stations) to monthly data. We present our
results of scaling analysis of volatility series obtained
from temperature data simulated by models, for
three different scenarios and
also for the observed data in the form of histograms e.g. Fig. 3.

The scaling exponents $\alpha$ obtained for the observed data are
shown in the topmost panel (for
comparison they are plotted repeatedly for three times corresponding to
three different scenarios). We can see immediately that all sites occupy
the bin corresponding to the $\alpha$ value in the range of 0.58 to 0.62 (see
Fig. 3 top 
panels) indicating the
universal behavior as found in the daily data. The grey
color boxes represent the $\alpha$ values of the volatility series derived
from surrogate series. For the case of monthly data the $\alpha$ values
obtained from the volatility series of surrogate data 
 exhibit correlated behavior for some of the sites. However, they are well below the values of the original
data. When we compare the results obtained from GCMs, none of the
models show a unique behavior as found in the observed data. The $\alpha$
values are distributed widely for all the models. The comparison of
the exponents of volatility series obtained from original
data and that of surrogate data shows their difference is close to
zero and even zero for some of the sites. For
instance, in CSIRO model, for each of the 
three different scenarios the exponent values of the volatility series
of the original data
and that derived from surrogate series, have same value for
St. Petersburg(9). Likewise, in ECHAM4, similar conclusion can be
drawn for Kasan(5) as well as Luling(6). However, the distribution
of $\alpha$ values of volatility series obtained from
surrogate data, towards higher values is clearly seen for the well
established scenarios like GHGF and GHGPS. 
\begin{figure}
\begin{center}
\includegraphics[width=4in,angle=0]{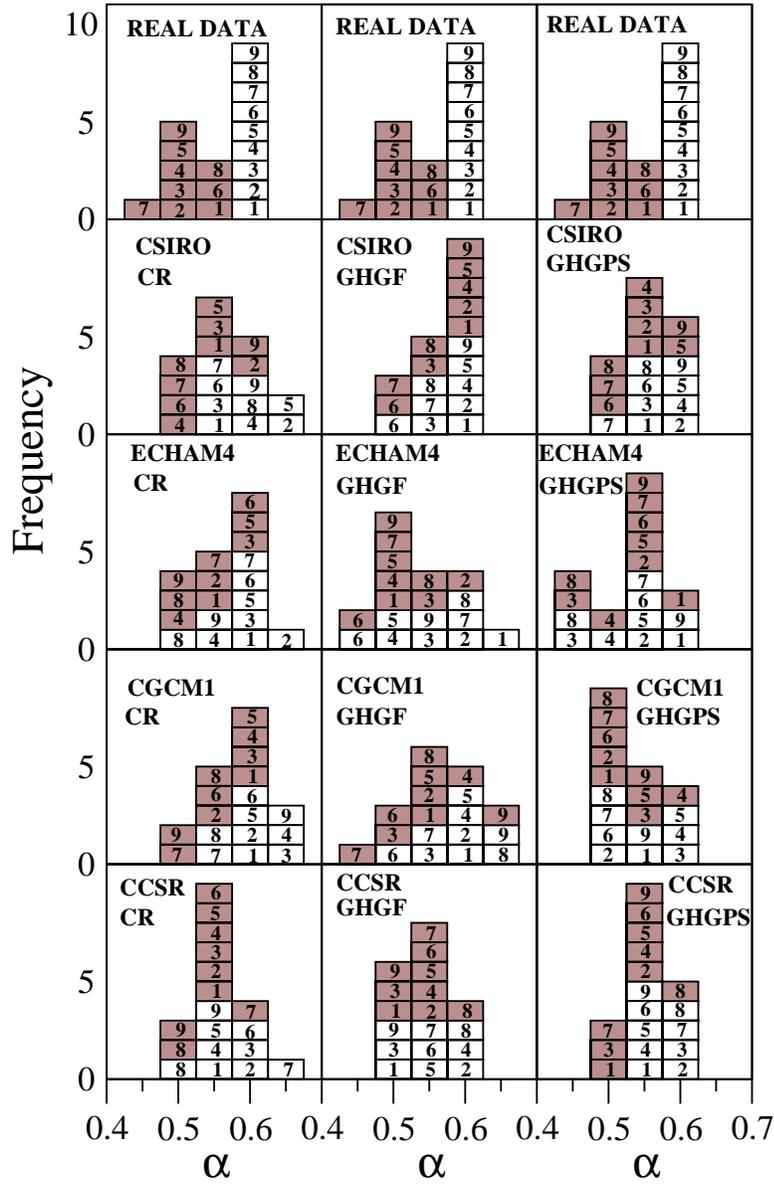} 
\caption{ 
Histogram of DFA exponent $\alpha$ for volatility series obtained from
monthly temperature data for nine different sites: Prague(1), Melbourne(2), Seoul(3), 
Vancouver(4),  Kasan(5), Luling(6), Tashkent(7), New York(8) and
St. Petersburg(9). Top 
three panels represent observed data. The other panels represent
the exponents of volatility series obtained from temperature time series simulated
by four different climate models (CSIRO-Mk2, ECHAM4/OPYC3, CGCM1 and
CCSR) for three different scenarios, namely control run (CR),
Greenhouse gas only forcing (GHGF) and Greenhouse gas plus sulphate
forcing (GHGPS). Open boxes represent the exponents obtained for
volatility series of original data  while grey boxes represent the exponents
obtained for volatility series derived from surrogate data of the
corresponding temperature increment series.
}
 
\label{fig3}
\end{center}
\end{figure}

It follows from our analysis that there is a universal persistence
in the volatility series obtained from temperature increment series
with an exponent of 
$0.60\pm0.03$. When we use this result to test the scaling performance
of the virtual climate records simulated by
GCMs, we find that (i) models data display wide range of exponent
values, (ii) surrogate analysis suggests that models
data lack nonlinearity, especially for the well established
scenarios, for some of the sites considered here.

\ack
This work has been supported by the Deutsche Forschungsgemeinschaft
and the Israel Science Foundation. We wish to thank Y. Ashkenazy for useful
discussions.


\begin{thebibliography}{}
\bibitem{6}
To eliminate climatological annual cycle from a series $x_i$, we
calculate the departures of $x_i$, $x_i - \langle x_i
\rangle$, from the mean values of $\langle x_i
\rangle$ for each 
calendar day $i$, say 1 January, which has been obtained by averaging
over all years in the series.
\bibitem{16}
Following \cite{4}, we define a monofractal signal as a linear signal and a multifractal
signal as a nonlinear signal. Volatility analysis (scaling analysis of the
magnitude series) of a monofractal signal with $\alpha$
values between $0.5\le \alpha \le 1.5$  will yield an uncorrelated
behavior. Volatility analysis of the 
surrogate data of a monofractal signal will yield uncorrelated behavior. In contrary, 
volatility analysis of a multifractal signal will yield correlated
behavior. Further, the volatility analysis of the
surrogate data of the multifractal signal will yield uncorrelated
behavior. Thus, for a multifractal signal, 
there exists a significance of difference between the volatility exponents of
the data and its (volatility series derived from) surrogates.
\end{thebibliography}
\end{document}